\documentclass[twocolumn]{aastex701}

\usepackage{amsmath}
\usepackage{cleveref}
\usepackage{array}
\usepackage{multirow}
\usepackage{soul}

\crefformat{figure}{Fig.~#2#1#3}
\crefformat{equation}{equation~(#2#1#3)}
\crefformat{section}{section~#2#1#3}
\crefformat{table}{Tab.~#2#1#3}
\crefformat{appendix}{appendix~#2#1#3}
\crefmultiformat{figure}{Figs.~#2#1#3}{~and~#2#1#3}%
    {,~#2#1#3}{,~#2#1#3}
\crefrangeformat{figure}{Figs.~(#3#1#4--#5#2#6)}
\crefrangeformat{equation}{equation~(#3#1#4--#5#2#6)}

\begin{document}

\title{The magnitude of the dark ages 21-cm signal in the context of existing early and late time constraints on $\Lambda$CDM}

\author[sname='Harry Bevins']{Harry T. J. Bevins}
\affiliation{Kavli Institute of Cosmology Cambridge, University of Cambridge,
Madingley Road,
Cambridge
CB3 0HA}
\affiliation{Cavendish Astrophysics, Cavendish Laboratory
JJ Thomson Avenue
Cambridge CB3 0HE}
\email[show]{htjb2@cam.ac.uk}  

\begin{abstract}
The dark ages 21-cm signal is a promising probe of the currently unobserved infant universe between the formation of the Cosmic Microwave Background around $z \approx 1100$ and the first galaxies around $z\approx 30$. A detection of the signal will help researchers understanding the nature of dark matter and dark energy, the expansion of the universe and any extensions to the concordance $\Lambda$CDM model that could explain the reported Cosmic Dawn 21-cm signal from EDGES and the Hubble tension. In this letter we take existing constraints on the $\Lambda$CDM cosmological model from two early time probes, Planck and WMAP, and two late time probes, DES galaxy lensing and clustering and Baryon Acoustic Oscillations, and propagate these through to constraints on the magnitude of the dark ages 21-cm signal. We constrain the magnitude and central frequency of the signal while methodically accounting for uncertainties in the cosmological parameters. We find that within the context of our modelling assumptions and the $\Lambda$CDM paradigm, the depth of the dark ages 21-cm signal is known to  better than 1 mK and the central frequency to within 0.05 MHz.
\end{abstract}



\section{Introduction}

The period in cosmic history between the formation of the Cosmic Microwave Background and the present Epoch of the Galaxies is largely unobserved. It covers the Dark Ages before the first galaxies formed, the Cosmic Dawn when the first galaxies and stars formed and the Epoch of Reionization when the universe transitioned from completely neutral to a near perfectly ionized state. Collectively we often refer to this period as the infant universe and while JWST has given us a handful of observations of high redshift galaxies up to $z \approx 14.4$ \citep{Naidu2025MoM-z14} this period between redshifts $1100$ and $10$ is largely unobserved. 

A promising probe of this epoch is the 21-cm signal from neutral hydrogen in the early universe \citep{Furlanetto200621cmReview, Barkana2016Review, Mesinger2019Review}. The signal arises from the spin-flip transition in which the electron's spin spontaneously flips from aligned to anti-aligned with the proton's spin, emitting a photon with a wavelength of 21-cm. At different periods during the infant universe the transition is mediated by photons from the CMB, collisional interactions between hydrogen atoms and light from the first stars and galaxies. By tracing the relative number of hydrogen atoms in each state, we can therefore learn about the evolution of the very early universe.

A number of experiments are trying to observe the sky-averaged signal from the cosmic dawn including REACH \citep{Acedo2022REACH}, EDGES \citep{Bowman2018EDGES}, SARAS \citep{Singh2022SARAS3, Bevins2022SARAS3} and MIST \citep{Monsalve2024MIST} among others as well as spatial fluctuations in the signal including HERA \citep{deboer2017HERA, HERA2023}, MWA \citep[e.g.][]{Nunhokee2025MWA, Trott2025MWA} and LOFAR \citep[e.g.][]{Acharya2024Lofar, Ceccotti2025lofar}. The EDGES collaboration reported a tentative detection of the sky-averaged cosmic dawn 21-cm signal in 2018 \citep{Bowman2018EDGES}, although the cosmological nature of this signal has been disputed in the literature \citep[e.g.][]{Hills2018EDGES, Sims2020EDGES, Singh2019EDGES, Bevins2021maxsmooth} and recent results from the SARAS3 instrument appear to be in tension \citep{Singh2022SARAS3}.

In order to observe the dark ages 21-cm signal, which is a rich probe of cosmology between $z=1100$ and $z=30$, researchers have to get above the ionosphere, which strongly attenuates the radio sky below 40 MHz. A number of experiments have been proposed to observe the dark ages and cosmic dawn 21-cm signal from the radio quiet lunar farside including CosmoCube \citep{Zhu2025CosmoCube, Artuc2025CosmoCube}, the Lunar Crater Radio Telescope \citep{Goel2022LCRT} and FarView \citep{Polidan2024FarView} among others. For a review of proposed missions and prospects for dark ages observations from the moon see \cite{Burns2021review}.

The reported EDGES absorption feature if cosmological in nature requires a much cooler gas temperature or stronger radio background than predicted in standard $\Lambda$CDM scenarios. Excess radio backgrounds from primordial black holes, early galaxies or cosmic strings \citep{Feng2018RB, Jana2019RB,Reis2020RB, Fialkov2019RB, Mirocha2019RB, EwallWice2018RB, GesseyJones2024RB} and interactions between baryonic matter and dark matter \citep{Barkana2018PossibleIB, Fialkov2018baryondm, Liu2019baryondm, Munoz2018baryondm, Berlin2018baryondm, Kovetz2018baryondm,Barkana2018baryondm, Slatyer2018baryondm} have all been proposed to explain the detection. Furthermore, a discrepancy has arisen between measurements of the expansion rate of the universe, known as the Hubble constant, from early time (CMB) and late time (Type Ia supernova) probes of cosmology \citep[see][]{Freedman2021Tensions}. An observation of the dark ages absorption feature could help researchers pin down the nature of dark matter \citep{Mondal2024darkagesdarkmatter} and dark energy, if cosmological in nature resolve the discrepant measurements of Hubble constant and explaining the anomalous depth of the EDGES absorption feature.

However, in the absence of physics beyond the standard $\Lambda$CDM cosmology, the dark ages 21-cm signal can act as probe of the matter density and dark energy density, the expansion rate of the universe and the composition of the gas. The signal can break degeneracies between the primordial helium fraction $Y_{\rm He}$ and baryon density $\Omega_b$ that are present in CMB-only analyses. Furthermore, when combined with low-redshift BAO measurements the signal can serve as a high-redshift anchor for the expansion rate, allowing independent constraints on $\Omega_m$ and $H_0$—analogous to how BBN is traditionally used. This makes the dark ages 21-cm signal a powerful complementary probe.

The standard $\Lambda$CDM cosmological model is already well constrained by existing probes like Planck, the Dark Energy Survey~(DES) and observations of Baryon Acoustic Oscillations~(BAO).  The relative constraining power of the dark ages 21-cm signal compared to Planck was explored in \cite{Mondal2023prospects}. Furthermore, existing works have looked at how the Maximum a Posteriori constraint on the cosmological parameters from Planck propagate through to the dark ages 21-cm signal and some have looked at the range of signals allowed given the mean and one sigma error bars on the parameters from Planck and other probes \citep[e.g.][]{Mittal2025echo21}. In this letter, we constrain the  magnitude of the dark ages 21-cm signal with existing early and late time cosmological probes by propagating existing MCMC chains through a model of the 21-cm signal to recover the functional posterior. By working directly with the MCMC chains from Planck and other probes we correctly propagate uncertainties and degeneracies in the cosmological parameters through to constraints on the magnitude of the 21-cm signal. We briefly discuss the implications for observations of the dark ages signal.

In Section \ref{sec:modelling} we outline our model for the 21-cm signal. In Section \ref{sec:existing_constraints} we introduce the existing cosmological probes and MCMC chains used in this analysis and in Section \ref{sec:magnitude} we present the corresponding functional posteriors on the dark ages 21-cm signal. We conclude in Section \ref{sec:conclusions}. The code used in this paper is publicly available at \url{https://github.com/htjb/existing-probes-dark-ages} and is archived at \url{https://doi.org/10.5281/zenodo.17910173}.

\section{Modelling the Dark Ages 21-cm Signal}
\label{sec:modelling}

Below we briefly discuss how we model the dark ages 21-cm signal and we refer the reader to the reviews by \cite{Furlanetto200621cmReview, Barkana2016Review, Mesinger2019Review} for more details. We are interested in the sky-averaged 21-cm signal in this work and consequently use a simple analytic model for the average evolution over cosmic time. More detailed semi-numerical models that evolve coupled differential equations for the density fluctuations, gas temperature, neutral fraction and the 21-cm signal in pixelated models of the Universe have been explored in \cite{Mondal2023prospects, Mondal2024darkagesdarkmatter, Flitter202421cmFirstClass}.

We quantify the number of atoms in each spin state with a statistical temperature known as the spin temperature $T_s$
\begin{equation}
    \frac{n_1}{n_0} = 3 \exp\bigg(\mathbf{ -}\frac{T_*}{T_s}\bigg),
\end{equation}
where $T_*$ is the equivalent temperature of a 21-cm photon and $n_1$ and $n_0$ are the number of atoms in each state.

During the early evolution of the universe, the relative number of neutral hydrogen atoms in each spin state is driven by interactions with CMB photons with a wavelength of 21-cm. In the absence of any other physics, the spin temperature would remain coupled to the CMB as it adiabatically cools. However, collisions between different hydrogen atoms also cause spin-slip transitions and couple the spin temperature to the gas kinetic temperature $T_k$. The strength of the collisional coupling is given by $x_c$ and the strength of coupling to the radio background or CMB $T_\gamma$ is given by $x_\gamma$. $x_\gamma$ is usually assumed to be 1 and so the spin temperature is given by
\begin{equation}
    T_s^{-1} = \frac{T_\gamma^{-1} + x_c T_k^{-1}}{1 + x_c}.
\end{equation}
We measure the spin temperature relative to the radio background 
\begin{equation}
    T_{21} = \frac{T_s - T_\gamma}{1+z}~(1 - \exp(-\tau_{21})) \approx \frac{T_s - T_\gamma}{1+z}~\tau_{21},
\end{equation}
and because the gas cools quicker due to adiabatic expansion than the CMB, we expect the 21-cm signal to appear in absorption. $\tau_{21}$ is the optical depth to the CMB and the approximation in the above holds in the optically thin limit. Assuming a homogenous IGM and expanding out $\tau_{21}$ we can write the 21-cm brightness temperature as
\begin{equation}
    \begin{aligned}
        T_{21} = 54 (1 - x_e) &\frac{1- Y_\mathrm{He}}{0.76} \frac{\Omega_b h^2}{0.02242} \\ & \sqrt{\frac{0.1424}{\Omega_m h^2} \frac{1+z}{40}} \bigg(1 - \frac{T_\gamma}{T_s}\bigg) {\rm [mK]},
    \end{aligned}
\end{equation}
where $x_e$ is the free electron fraction, $Y_\mathrm{He}$ is the helium fraction, $\Omega_b$ and $\Omega_m$ are the baryon and matter densities, $h$ is reduced Hubble's constant $H_0$ and $z$ is redshift \citep{Mondal2024darkagesdarkmatter}.

The collisional coupling coefficient $x_c$ is given by
\begin{equation}
    x_c =  \frac{T_*}{A_{10} T_\gamma} n_H (\kappa_\mathrm{HH} (1 - x_e) + (\kappa_\mathrm{eH} + \kappa_\mathrm{pH}) x_e),
    \label{eq:collisional-coupling}
\end{equation}
where $A_{10}$ is the Einstein coefficient for the transition between each spin state. $\kappa_\mathrm{HH}$, $\kappa_\mathrm{eH}$ and $\kappa_\mathrm{pH}$ are the rate coefficients for Hydrogen-Hydrogen, electron-Hydrogen and proton-Hydrogen collisions. During the dark ages when $x_e$ is low, collisions between hydrogen atoms dominates over collisions between hydrogen and free electrons or hydrogen and free protons. We assume $\kappa_\mathrm{eH}$ and $\kappa_\mathrm{pH}$ are roughly equal. In practice, they differ by a factor related to the ratio of the proton and electron masses, but since they are subdominant at the relevant redshifts, this assumption makes little difference. The collisional rate coefficients can be approximated as
\begin{equation}
    \begin{aligned}
        \kappa_{HH} = & 3.1 \times 10^{-17}~T_k^{0.357}~\exp\bigg(-\frac{32}{T_k}\bigg), \\
        \log \kappa_{eH} = & \log \kappa_{pH} = -9.607 + \\ & 0.5 \log T_K \exp \bigg[ - \frac{(\log T_k)^{4.5}}{1800}\bigg],
    \end{aligned}
    \label{eq:kappa}
\end{equation}
in m$^3$s$^{-1}$ where $n_H$ is the number density of neutral hydrogen atoms
\begin{equation}
    n_H = \frac{(1 - Y_\mathrm{He})}{m_p} \frac{3 H_0^2}{8 \pi G} \Omega_b (1 + z)^3.
\end{equation}
$G$ is Newtons constant of gravity, $m_p$ is the proton mass and $H_0$ is Hubble's constant \citep{Moazzenzadeh2021kappas}.

In order to evaluate the sky-averaged 21-cm signal during the dark ages, we need the electron number density $x_e$ and the gas temperature $T_k$ as functions of redshift. We obtain these quantities using \texttt{HYREC-2} \citep{Nanoom2020Hyrec2} and separately with \texttt{recfast++} \citep{Chluba2010Recfast++, Chluba2011Recfast++}. Both of these codes compute the recombination history of hydrogen and helium in the early universe, however they use slightly different approaches. \texttt{recfast++} uses a simplified 3 level atom model, whereas \texttt{HYREC-2} accounts for a virtually infinite number of energy levels in the atom and uses full radiative transfer calculation for important transitions. We find that the differences in the calculated $x_e$ and $T_k$ are essentially negligible. 
We also note that existing semi-numerical frameworks for modelling the evolution of the dark ages 21-cm signal like 21cmFirstCLASS typically use \texttt{HYREC-2} \citep{Flitter202421cmFirstClass} to set the initial values of $x_e$ and $T_k$ but then evolve these quantities in each pixel of the simulation via coupled differential equations with the density fluctuations. 


Our model is a function of the cosmological parameters $\Omega_m$, $\Omega_b$, $\Omega_c$, $H_0$ and $Y_\mathrm{He}$.

\section{Existing Constraints}
\label{sec:existing_constraints}

In this paper, we investigate the constraining power of the Planck CMB high l TT + low l + low E likelihood \citep{Planck2020}, the WMAP full CMB (polarisation and temperature) 9 year likelihood \citep{Hinshaw2013WMAP}, combined BAO measurements from the DR12 \citep{Alam2017DR12}, MGS \citep{Ross2015MGS} and 6dF \citep{Beutler20116DF} catalogues and DES Y1 cosmic shear, galaxy clustering and cross correlation~\citep{Abbott2018DESY1} on the dark ages signal. MCMC chains from fits of $\Lambda$CDM to these data sets are publicly available from the Planck Legacy Archive~(PLA)\footnote{\url{https://pla.esac.esa.int/}} and it is these that we use to understand the existing constraints on the dark ages 21-cm signal. We note that newer constraints from ACT \citep[e.g.][]{Louis2025Act, ACT2024Pol}, DESI \citep[e.g.][]{Adame2024DESIa, Adame2025Desi}, DES \citep[e.g.][]{Abbott2022DES} and other probes exist but we leave exploration of these for future work and continue with the four results from the PLA.

While not all the parameters of interest ($H_0$, $\Omega_m$, $\Omega_b$, $\Omega_c$ and $Y_\mathrm{He}$) are directly probed by the different data sets the PLA chains include derived parameters. For example, $Y_\mathrm{He}$ is calculated from $\Omega_b$ using the Parthenope 1.1 BBN code and $\Omega_m$ is calculated from $\Omega_b h^2$ and $\Omega_c h^2$. For Planck and WMAP $H_0$ is also derived from constraints on other parameters such as the angular scale of the sound horizon at decoupling. We refer the reader to \cite{Planck2020} and the PLA\footnote{\url{https://wiki.cosmos.esa.int/planck-legacy-archive/index.php?title=Cosmological_Parameters}} for more details on the chains.

\section{The Magnitude of the Dark Ages 21-cm signal}
\label{sec:magnitude}

To demonstrate the constraining power of the different cosmological probes explored in this work, we define a realistic prior on the cosmological parameters probed by the dark ages 21-cm signal (see \cref{tab:prior}). It is not the prior used by the Planck collaboration, but is representative of the prior that could be used in the analysis of data from a lunar 21-cm experiment.

\begin{table}[]
    \centering
    \begin{tabular}{|c|c|}
    \hline
        Parameter & Prior \\
        \hline
         $H_0$ & $\mathcal{U}(40, 100)$\\
         $\Omega_m$ & $\mathcal{U}(0.1, 0.5)$ \\ 
         $\Omega_b$ & $\mathcal{U}(0.005, 0.1)$\\ 
         $\Omega_c$ & $\Omega_m - \Omega_b$\\
         $Y_\mathrm{He}$ & $\mathcal{U}(0.2, 0.3)$\\
    \hline
    \end{tabular}
    \caption{A representative prior on the cosmological parameters. This is used to derive a representative functional prior on the dark ages 21-cm signal. We note that this is not the prior used in the MCMC analysis of Planck, WMAP, DES Y1 or the BAO data.}
    \label{tab:prior}
\end{table}

\begin{figure*}
    \centering
    \includegraphics[width=\linewidth]{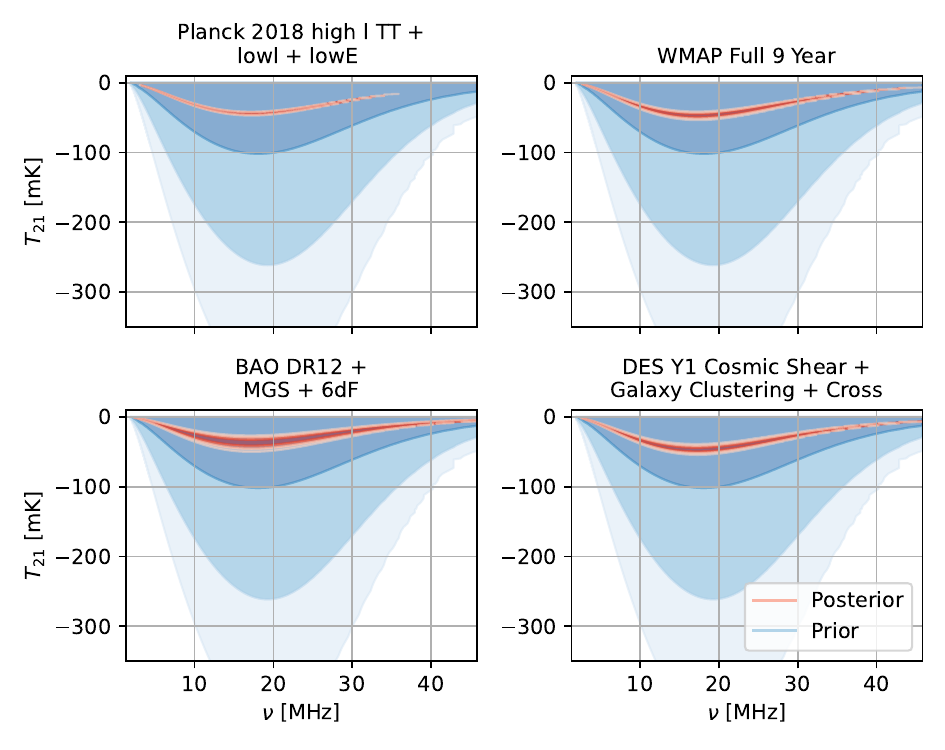}
    \caption{The functional posteriors (red) on the dark ages 21-cm signal from constraints on the parameters of $\Lambda$CDM cosmology from four existing probes using the \texttt{HYREC-2} recombination code; Planck and WMAP observations of the CMB, BAO observations and lensing and galaxy clustering from DES. A representative prior is shown in blue. The darker shaded regions correspond to 1 sigma contours, with the lighter shaded regions being 2 and 3 sigma. The figure is made with \texttt{fgivenx} \citep{Handley2018fgivenx}.}
    \label{fig:signals-hyrec}
\end{figure*}

\begin{figure}
    \centering
    \includegraphics[width=\linewidth]{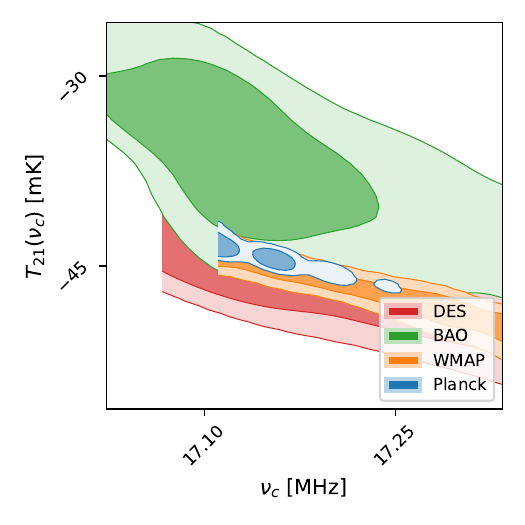}
    \caption{The posterior on the central frequency and depth of the dark ages 21-cm signal from the four cosmological probes explored in this work using \texttt{HYREC-2}. The  figure was produced with \texttt{anesthetic} \citep{Handley2019anesthetic}.}
    \label{fig:minima-hyrec}
\end{figure}

\Cref{fig:signals-hyrec} shows the functional posteriors for the dark ages 21-cm signal from Planck and WMAP CMB data, BAO observations from SDSS and DES galaxy clustering and lensing constraints on the cosmological parameters using \texttt{HYREC-2}. We also show in each panel the functional prior corresponding to the parameter prior detailed in \cref{tab:prior}. The frequency range corresponds to redshifts 1100 to 30. \cref{fig:minima-hyrec} shows the posteriors on the central frequency and minimum temperature of the sky-averaged 21-cm calculated using \texttt{HYREC-2}. In \cref{tab:minima} we show the mean and standard deviation on these values calculated with both \texttt{HYREC-2} and \texttt{recfast++}.

\begin{table*}[]
    \centering
    \begin{tabular}{|c|c|c|c|c|}
    \hline
    Data Set & \multicolumn{2}{c|}{\texttt{recfast++}}& \multicolumn{2}{c|}{\texttt{HYREC-2}} \\
    \cline{2-5}
    & $T_{21}(\nu_c)$ [mK] & $\nu_c$ [MHz] & $T_{21}(\nu_c)$ [mK] & $\nu_c$ [MHz]\\
    \hline
    Planck 2018 high $l$ TT + lowl + lowE & $-44.37 \pm 0.96$ & $17.14 \pm 0.02$ &  $-44.09 \pm 0.96$ & $17.14 \pm 0.03$ \\ 
    WMAP Full 9 Year & $-47.35 \pm 2.10$ & $17.20 \pm 0.06$ &  $-47.04 \pm 2.08$ & $17.23 \pm 0.07$  \\
    BAO DR12 + MGS + 6dF & $-37.32 \pm 4.10$ & $17.14 \pm 0.04$ & $-37.08 \pm 4.07$ & $17.14 \pm 0.05$  \\ 
    DES Y1 Cosmic Shear + Galaxy Clustering + Cross &  $-46.89 \pm 2.55$& $17.16 \pm 0.05$ & $-46.58 \pm 2.53$ & $17.18 \pm 0.06$ \\
    Prior &  $-82.08 \pm 91.08$ & $17.04 \pm 1.63$ & $-82.36 \pm 89.73$  & $17.52 \pm 3.72$  \\
    \hline
    \end{tabular}
    \caption{The mean and standard deviations of the 1D posteriors on the central frequency and corresponding value of the 21-cm temperature from the four probes of $\Lambda$CDM cosmology explored in this paper using both \texttt{recfast++} and \texttt{HYREC-2}. The corresponding values for the representative prior are also shown.}
    \label{tab:minima}
\end{table*}

\begin{figure*}
    \centering
    \includegraphics[width=0.8\linewidth]{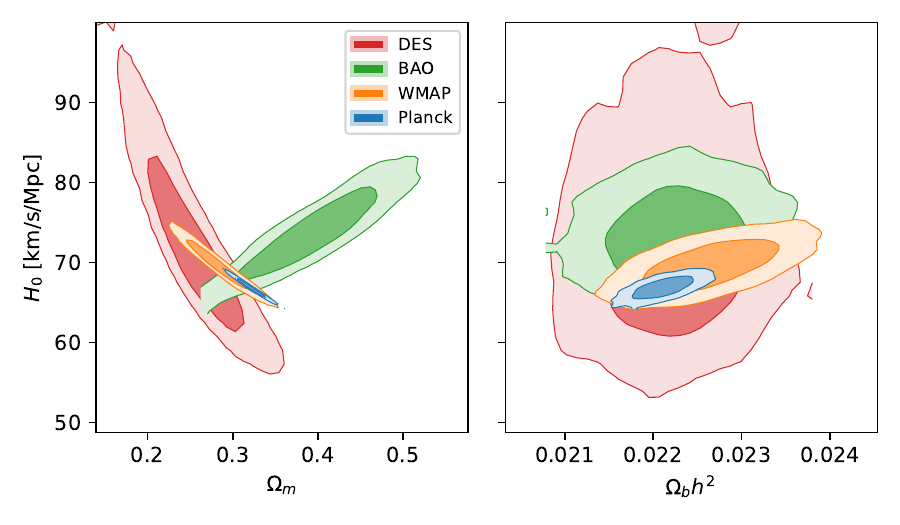}
    \caption{The posterior constraints on the present day expansion rate of the universe $H_0$, the total baryon density $\Omega_b$ and the total matter density $\Omega_m$ from the four data sets explored in this work.}
    \label{fig:parameter_constraints}
\end{figure*}

The magnitude and timing of the dark ages 21-cm signal is strongly dependent on the present day expansion rate of the universe $H_0$, the baryon density $\Omega_b$ and the total matter density of the universe $\Omega_m$. We can therefore understand the constraints presented in \cref{fig:signals-hyrec} and \cref{fig:minima-hyrec} by looking at the constraints on these parameters from the four different cosmological probes, shown in \cref{fig:parameter_constraints}.

The first thing to note is that the constraints on these parameters from the two late time probes, DES Y1 and BAO observations, are weaker than those from the CMB observations. This means that the uncertainty on the depth and central frequency of the dark ages 21-cm signal from these probes is larger.

The second thing to consider is that the value of $\Omega_b$ modulates the rate of collisional coupling. A higher value leads to stronger collisional coupling between the spin temperature and gas temperature and a deeper absorption signal. However, all four probes constrain $\Omega_b$ to very similar values and so any differences in the predicted signals are from other parameters.

A higher matter density and a higher Hubble parameter both lead to a faster expansion of the universe. A faster expansion leads to an earlier decoupling of the gas from the CMB and although the gas then cools adiabatically quicker than the CMB it starts cooling from a higher temperature, meaning the gas is warmer at the redshifts relevant for the dark ages 21-cm signal. Furthermore, a faster expansion rate means that collisional coupling becomes inefficient more quickly and the spin temperature decouples from the gas at earlier times. As a result, the 21-cm signal is shallower for larger values of $H_0$ and $\Omega_m$.

The BAO observations favour a higher matter density than the other probes, meaning that they favour shallower absorption features. While DES permits larger values of $H_0$ than the other probes, the $H_0$--$\Omega_b$ posterior exhibits a negative correlation, such that increases in $H_0$ and decreases in $\Omega_b$ act in opposite directions on the signal amplitude. Overall, the constraints on the depth of the dark ages 21-cm signal from these four probes are consistent within roughly 2$\sigma$.

The observations from Planck indicate that the dark ages 21-cm signal, in the $\Lambda$CDM cosmology using the \texttt{HYREC-2} recombination code, has a central frequency of $17.14 \pm 0.03$ MHz and a depth equal to $-44.09 \pm 0.96$ mK.

\section{Conclusions}
\label{sec:conclusions}

The dark ages 21-cm signal is a potentially powerful probe of a previously unseen epoch in cosmic history. In this paper, we have built a simple analytic model of the sky-averaged signal using both \texttt{HYREC-2} and \texttt{recfast++} to model the evolution of the intergalactic medium. We then demonstrated that existing constraints from early and late time probes on the $\Lambda$CDM cosmological model can be used to infer the magnitude of the signal.

New data from DES, ACT and DESI may change the results presented here marginally, but we leave an investigation of the constraining power of these probes to future work. We have also not explored the constraining power of different Type Ia supernova catalogues (e.g. Pantheon+ \cite{Brout2022Pantheon}, DES5Y \cite{DESY52024}, SH0ES \cite{Riess2022Sh0es}) on the magnitude of the dark ages 21-cm signal. Current compilations of Type Ia supernova exhibit mild but non-negligible tensions and calibration differences with each other, and were therefore not included here to reduce systematic complexity and uncertainty.

While the differences between predictions computed with \texttt{HYREC-2} and \texttt{recfast++} are well below the $\Lambda$CDM-driven uncertainty considered here, they may become relevant once dark-ages measurements reach substantially higher precision. In principle, one can assess such modelling choices using Bayesian evidence or by marginalising over them, treating the recombination history as a source of theoretical uncertainty. Looking ahead, combining independent probes—CMB, large-scale structure, and eventually dark-ages 21-cm measurements—offers a way to break degeneracies and cross-check for consistency. Such synergies must, however, be explored with caution to avoid conflating genuinely incompatible datasets.

Detecting the 21-cm signal from the Cosmic Dawn, Epoch of Reionization and the Dark ages represents a significant data analysis challenge. While the discovery space is in extensions to the standard cosmological model, by studying the constraints from existing probes of $\Lambda$CDM we can start to understand the observational requirements for lunar missions. The results presented in this short paper suggest that future lunar missions attempting to detect the dark ages 21-cm signal should optimise their instruments to detect signals around 17 MHz or redshifts 85. It also sets stringent requirements on the calibration of observations if researchers want to be competitive with constraints from Planck. Most ground based instruments aiming to observe the cosmic dawn 21-cm signal are aiming to calibrate their data to around 25 mK \citep[e.g.][]{Roque2021Calibration, Murray2022Calibration, Roque2025Calibration, Kirkham2015Calibration, Leeney2015MLCalibration} or a signal-to-noise~(SNR) ratio of approximately 8 assuming a 21-cm signal of around 200 mK \citep[e.g.][]{Reis2021lymanalpha}. To achieve a similar SNR lunar missions need to calibrate to around $44/8 \approx 5.5$ mK and to be competitive with Planck and WMAP they need to calibrate to better than $1$ mK. While there have been significant recent advancements in the calibration of ground based instruments \citep[e.g.][]{Murray2022Calibration, Kirkham2015Calibration, Leeney2015MLCalibration}, larger foregrounds at lower frequencies, larger temperature swings in lunar orbit and restrictions on space make the calibration of dark ages observations even more daunting.

\begin{acknowledgments}

\end{acknowledgments}

HTJB acknowledges support from the Kavli Institute for
Cosmology Cambridge and the Kavli Foundation and Jiten Dhandha for useful conversations.

\begin{contribution}

All authors contributed equally to the manuscript.


\end{contribution}

%

\software{\texttt{HYREC-2}\citep{Nanoom2020Hyrec2},  
          \texttt{recfast++} \citep{Chluba2010Recfast++},
          \texttt{anesthetic} \citep{Handley2019anesthetic},
          \texttt{fgivenx} \citep{Handley2018fgivenx}
          }



\clearpage
\bibliography{sample701}{}
\bibliographystyle{aasjournalv7}



\end{document}